\documentclass[conference]{IEEEtran}
\IEEEoverridecommandlockouts
\usepackage{cite}
\usepackage{amsmath,amssymb,amsfonts}
\usepackage{algorithmic}
\usepackage{graphicx}
\usepackage{textcomp}
\usepackage{xcolor}
\usepackage{array}
\usepackage{xspace}
\usepackage{multirow}
\usepackage[normalem]{ulem}
\useunder{\uline}{\ul}{}
\def\BibTeX{{\rm B\kern-.05em{\sc i\kern-.025em b}\kern-.08em
    T\kern-.1667em\lower.7ex\hbox{E}\kern-.125emX}}
\begin{document}

\title{Self-supervised Complex Network \\ for Machine Sound Anomaly Detection
}

\author{\IEEEauthorblockN{Miseul Kim}
\IEEEauthorblockA{\textit{DSP\&AI Lab., Dept. of E.E.} \\
Yonsei Univ., Seoul, Korea \\
miseul4345@dsp.yonsei.ac.kr}
\and
\IEEEauthorblockN{Minh Tri Ho}
\IEEEauthorblockA{\textit{DSP\&AI Lab., Dept. of E.E.} \\
Yonsei Univ., Seoul, Korea \\
mtho96@dsp.yonsei.ac.kr}
\and
\IEEEauthorblockN{Hong-Goo Kang}
\IEEEauthorblockA{\textit{DSP\&AI Lab., Dept. of E.E.} \\
Yonsei Univ., Seoul, Korea \\
hgkang@yonsei.ac.kr}
}

\maketitle

\begin{abstract}
In this paper, we propose an anomaly detection algorithm for machine sounds with a deep complex network trained by self-supervision.
Using the fact that phase continuity information is crucial for detecting abnormalities in time-series signals, our proposed algorithm utilizes the complex spectrum as an input and performs complex number arithmetic throughout the entire process. 
Since the usefulness of phase information can vary depending on the type of machine sound, we also apply an attention mechanism to 
control the weights of the complex and magnitude spectrum bottleneck features depending on the machine type.
We train our network to perform a self-supervised task that classifies the machine identifier (id) of normal input sounds among multiple classes.
At test time, an input signal is detected as anomalous if the trained model is unable to correctly classify the id. In other words, we determine the presence of an anomality when the output cross-entropy score of the multi-class identification task is lower than a pre-defined threshold. 
Experiments with the MIMII dataset show that the proposed algorithm has a much higher area under the curve (AUC) score than conventional magnitude spectrum-based algorithms.

\end{abstract}

\begin{IEEEkeywords}
complex-network, self-supervised classification, anomaly detection
\end{IEEEkeywords}

\section{Introduction}

Machine sound anomaly detection is the task of detecting abnormalities in a machine by analyzing the characteristic of its sounds \cite{unsupervised1,unsupervised2,unsupervised3}. 
Since malfunctioning machines in factories can have a significant impact on cost-effectiveness, it is important to detect anomalous signals to identify signs of failure as early as possible. 
The statistical properties of anomalous sounds are difficult to pre-define because anomalous sounds occur very infrequently in real situations.
Therefore, one way of solving this task is to use unsupervised or semi-supervised learning methods that utilize only normal data \cite{unsupervsied_ex}.

Unsupervised learning for anomaly detection can be divided into three categories: reconstruction, distribution, and feature learning based approaches \cite{category}.
Reconstruction based approaches use the difference between a reference input and a reconstructed output by assuming that the reconstruction accuracy of normal sounds is much higher than for abnormal sounds. 
Autoencoders \cite{autoencoder1,autoencoder2} and variational autoencoders~\cite{VAE} are typical models belonging to this category.
In \cite{autoencoder_gan}, the authors utilize a generative adversarial network (GAN) \cite{GAN} in a reconstruction based framework. 
However, its performance often degrades because the model also accurately reconstructs anomalous signals as well as normal signals.
Distribution based methods detect anomalies by measuring the statistical similarity between the input sound and a pre-trained distribution of normal data.
Gaussian mixture models (GMMs) are an example of a representative approach \cite{GMM, GMM1,GMM2}.   
In classification based approaches, representative feature embeddings are obtained by transforming input data into a feature domain;
then, classification tasks are carried out using these learned features. 
Since the embeddings include key characteristics of input data, it is much easier to determine anomalies using normal data.
An example of this type of approach is one-class support vector machines (OC-SVMs) \cite{OC-SVM}.

Recently, self-supervised representation learning approaches that do not need handcrafted labels have been used for anomalous sound detection \cite{dcase2020_outlier-exposer, dcase2020_giri, dcase2020_classification_confidence}.
In these approaches, some portion of the input is regarded as a target, which the model is trained to predict with the remaining portion of the data.
In \cite{dcase2020_classification_confidence}, two classifiers are used: one identifies machine type and id, and the other predicts the type of data augmentation.
An ensemble of masked autoencoders for density estimation (G-MADE) \cite{MADE} with linearly combined augmentation has also been proposed at \cite{dcase2020_giri}. 

Most conventional sound anomaly detection methods utilize magnitude spectra for input features; however, phase continuity information is also effective for a wide variety of sound types \cite{minh_multichannel, importance_phase, dcunet}. 
\cite{dcase2020_recognition} obtained high area under the curve (AUC) scores by introducing phase terms. However, since the phase terms were selectively used only for specific machine types (e.g. \textit{fan}), 
it is hard to generalize the method to be applicable for all machine types.


To resolve the limitations on generality, we propose an anomaly detection algorithm based on self-supervision with a deep complex network that utilizes an attention mechanism to control the weights for the magnitude and phase terms of the signal for different machine types.
Since the proposed network only learns normal machine sound information, anomalies can be detected when the shape of the output logits deviate significantly from a one-hot vector for the machine id type. 

Our paper is organized as follows. In Section \ref{sec:baseline}, we introduce a baseline model and describe its limitations. In Section \ref{sec:proposed}, we introduce our proposed model in detail. In Section \ref{sec:experiments}, we show experimental results that verify the superior performance of our model. Finally, Section \ref{sec:conclusion} provides our conclusions. 

\section{baseline model and limitations} 
\label{sec:baseline}
\subsection{Self-supervised classification with augmentation network}
\begin{figure}[t]
\centering
{\includegraphics[width=0.51\textwidth]{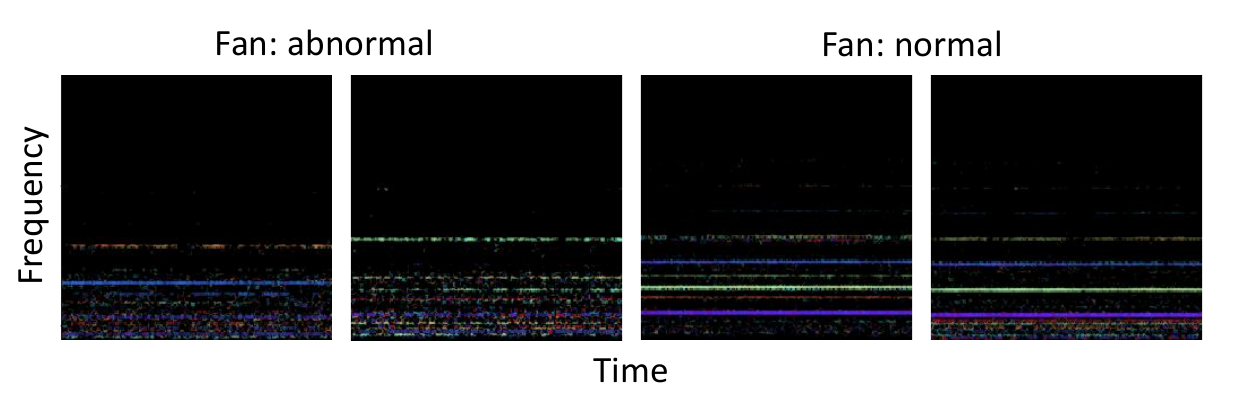}}
\hfill
{\includegraphics[width=0.51\textwidth]{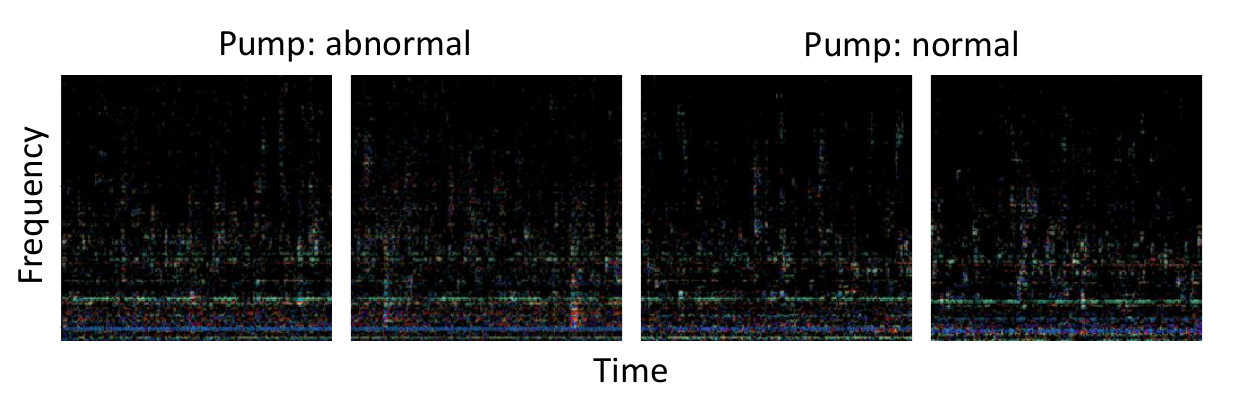}}
\caption{Rainbowgram of normal and abnormal sounds (upper row: id-06 of \textit{fan} sound, lower row: id-00 of \textit{pump} sound, both from the MIMII dataset \cite{mimii}). Left two columns show anomalous samples and right two columns show normal samples.}
\label{rainbow}
\end{figure}

As the baseline for our proposed model, we adopt the self-supervised classification network with linear combination augmentation \cite{dcase2020_giri} that showed very high performance in the DCASE2020 challenge \cite{dcase2020_task2}.
The baseline network consists of an ensemble of density estimation, group masked autoencoder \cite{MADE}, and a self-supervised classification module.
A decision on anomaly detection is made by the classification module, which is trained to identify the id of each machine type. 
The classification module is trained by a self-supervised learning technique with data augmentation.
Specifically, input log-mel spectrograms are artificially generated by linearly combining the log-mel spectrogram of each machine id with randomized weights. For example, given the log-mel spectrogram of machine id $i$, $x_i$, an augmented input data $x_n$ is obtained by a weighted sum of $x_i$, $\sum_{i=1}^{I} \alpha_i x_i$, where $I$ is the total number of machine ids.
The weights $\alpha_i$, which are determined randomly, sum to 1.  
The network parameters are trained to minimize the Kullback–Leibler (KL) divergence between the target reference $\alpha_1, \alpha_2, ..., \alpha_I$ and output logits.
At test time, the cross-entropy between the output logits of the trained model and the ground truth label represented by a one-hot vector is computed to determine the degree of abnormality.

\subsection{Phase continuity in anomaly detection}
Although the baseline model mentioned above shows good performance, the accuracy of anomaly detection is not high enough for some machine types \cite{dcase2020_giri}. 
The cause for this performance degradation can be found from the type of input features, log-mel spectrograms.
Log-mel spectrograms represent the mel-scaled magnitude of frequency bands to simulate the property of the human auditory system \cite{mel}. However, since anomaly detection must be regarded as a pattern matching process that is not related to perception, we may lose key information from the mel-scale representations. 
In order to further improve the anomaly detection accuracy of time-series signals, it is important to also utilize phase-related information such as phase continuity \cite{rainbowgram}.

Fig. \ref{rainbow} shows rainbowgrams \cite{rainbowgram} that display both the magnitude and phase spectra with a constant-q transform (CQT) \cite{CQT}. The magnitude spectrum is represented by line intensities and the derivative of phase spectrum is represented by colors.
For example, the rainbowgrams of normal fan sounds, shown in the third and fourth columns of the upper row, clearly display intense and identically colored continuous lines for each frequency bin. However, for abnormal sounds, shown in the first and second columns, both the thickness and colors are significantly different from the normal cases. 
However, for pump sounds, shown in the lower row, we are not able to clearly see the difference between normal and anomalous sounds using only the phase discontinuity information. Therefore, we expect that the effectiveness of using phase continuity information for anomalous machine sound detection varies by the type of machine. 

\section{Proposed complex neural network model}
\label{sec:proposed} 
\begin{figure}[t]
\centerline{\includegraphics[scale=0.6]{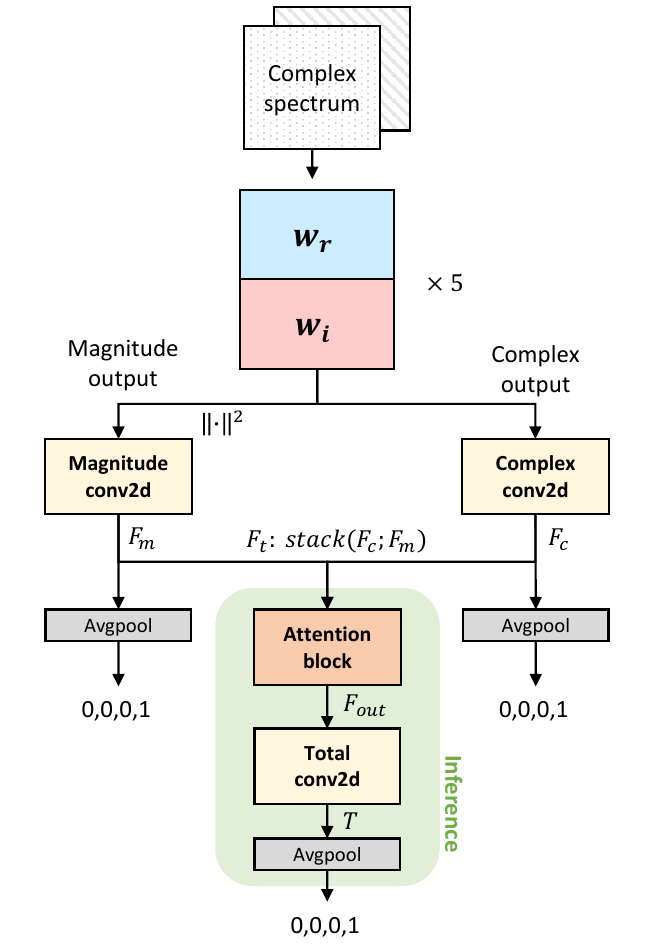}}
\caption{Structure of the proposed self-supervised complex network architecture. The red and blue colored box represents a real and imaginary part of complex convolutional block, respectively.}
\label{self_supervised_complex_network}
\end{figure}

Motivated by the structure of the above baseline model and the importance of phase continuity, we propose a complex spectrogram based anomaly sound detection model. 
Fig. \ref{self_supervised_complex_network} illustrates the overall structure of our proposed model.
We first calculate the complex spectra of the waveform input signals, then pass them to bottleneck blocks consisting of complex convolutional networks. We used an attention module to control the effect of phase for each machine type. After performing average pooling on the bottleneck output, we perform 3 machine identification tasks for each component. The model is trained with a sum of the three cross-entropy losses from these tasks.

\subsection{Self-supervised complex neural network}
The proposed self-supervised complex neural network performs a classification task to detect anomalous machine sounds. 
We use the encoder part of Deep Complex U-Net \cite{dcunet} to extract key features for identification, replacing Mobilenetv2 \cite{mobilenetv2}, which was used in the original baseline.
To capture phase-related information by keeping complex operations, the first 5 network layers are composed of a complex convolutional block \cite{dccrn}.
After each complex convolution block, batch normalization and a parametric ReLU activation function (PReLU) are applied for each complex component. Next, the complex output is divided into a complex component and magnitude component by taking its absolute value. The magnitude and complex components are fed through separate 2D-convolution layers, which output features $F_m$ and $F_c$. Then, $F_m$ and $F_c$ are recombined as $F_t\in \mathbb{R}^{2C\times F \times T}$ by stacking on the channel dimension. 

An attention mechanism is applied to $F_t$ to control the importance of phase continuity for a given machine type.
The output feature $F_{out}$, which is generated by attention block, is passed through another 2D convolution layer (named as "Total conv2d") in Fig. \ref{self_supervised_complex_network} and then feature $T$ is generated as an output. After that, average pooling is applied to all three sets of features, $F_m$, $F_c$ and $T$. Finally, the model performs three multi-class classification tasks with softmax outputs. The dimension of the output is determined by the number of classes. 

\subsection{Attention block}
\begin{figure}[t]
\centerline{\includegraphics[scale=0.48]{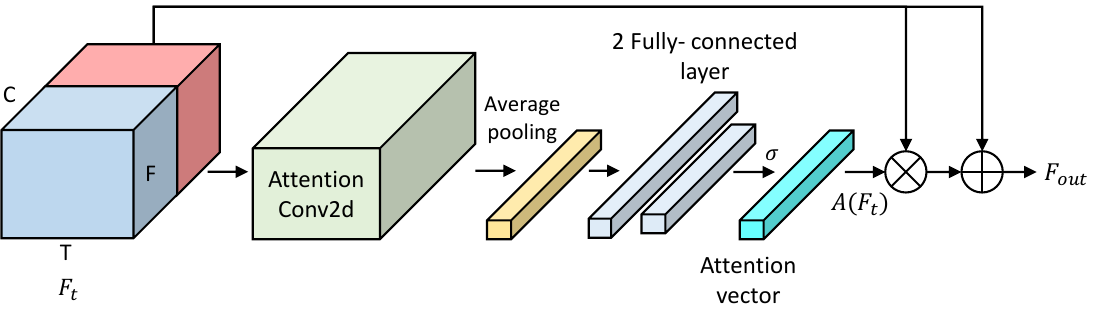}}
\caption{Structure of the attention block.}
\label{attention}
\end{figure}

As explained above, we used an attention module to control the relative importance of the magnitude and complex components for different machine types. 
Fig. \ref{attention} shows a specific structure to obtain the attention weight, $A$. 
The combined feature, $F_t$ are passed through a 2D convolution layers, then average pooling is applied to the output.
This is followed by 2 fully-connected (FC) layers. The first FC layer output dimension is 4 times as large as its input dimension, then it is recovered to the same dimension as the input at the second linear layer. Finally, the channel attention vector $A\in \mathbb{R}^{2C\times 1 \times 1}$ is generated by applying a softmax function, which is denoted as $\sigma$ in Fig. \ref{attention}. Finally Attention vector $A(F_t)$ is element-wise multiplied with $F_t$ and then residual connection is applied.
The process of the attention module can be represented as
\begin{equation}
\label{equation1}
F_{out} = A(F_t) \otimes F_t + F_t,
\end{equation}
where $A(\cdot)$ is an attention block and $\otimes$ denotes an element-wise multiplication. 
The structure of the convolution block is summarized in Table \ref{table0}.

\section{Experiments}
\label{sec:experiments}

\begin{table}[t]
\centering
\caption{Construction of convolution blocks}
\label{table0}
\begin{tabular}{|c|c|c|c|}
\hline
Operator         & Kernel size & Stride & Channel \\ \hline
Complex net1     & (7,5)       & (2,2)  & 45      \\ \hline
Complex net2     & (7,5)       & (2,2)  & 90      \\ \hline
Complex net3     & (5,3)       & (2,2)  & 90      \\ \hline
Complex net4     & (5,3)       & (2,2)  & 90      \\ \hline
Complex net5     & (5,2)       & (2,1)  & 45      \\ \hline
Magnitude conv2d & (5,1)       & (2,1)  & 4       \\ \hline
Complex conv2d   & (5,1)       & (2,1)  & 4       \\ \hline
Total conv2d     & (1,1)       & (1,1)  & 4       \\ \hline
Attention conv2d & (3,1)       & (2,1)  & 8       \\ \hline
\end{tabular}
\LARGE
\end{table}

\subsection{Datasets}
We used the MIMII dataset \cite{mimii} for the anomaly detection task, which was released by the DCASE2020 challenge. MIMII is a machine sound dataset encompassing normal and anomalous conditions that reflect real industrial situations. 
The anomaly sets include various anomalous conditions such as contamination, leakage, and rotation unbalance. Each sound is recorded as a single-channel 10-second segment and downsampled to 16kHz. The dataset consists of 4 machine types (fan, pump, valve, slider) and each machine type is divided into 4 machine ids. Specifically, the machine id denotes an identifier for each individual machine of the same machine type. Our proposed self-supervised classification model aims to classify these machine ids for each machine type. 


\begin{table*}[ht]
\centering
\caption{AUC scores obtained by self-supervised classification tasks with various setups}
\begin{tabular}{|c|c|c|c|c|c|c|c|}
\hline
\multirow{2}{*}{\begin{tabular}[c]{@{}c@{}}Self-supervised\\ classification\end{tabular}} & \multirow{2}{*}{\begin{tabular}[c]{@{}c@{}}Input\\ spectrum\end{tabular}} & \multirow{2}{*}{\begin{tabular}[c]{@{}c@{}}Network\\ architecture\end{tabular}}      & \multicolumn{4}{c|}{Machine type}                                 & \multirow{2}{*}{Avg(\%)} \\ \cline{4-7}
                                                                                          &                                                                           &                                                                                      & Fan            & Pump           & Slider         & Valve          &                          \\ \hline
\multirow{3}{*}{\begin{tabular}[c]{@{}c@{}}Giri, Ritwik et al. \\ {[}16{]}\end{tabular}}  & \multirow{2}{*}{Log-mel}                                                  & Mobilenetv2                                                                          & 80.61          & 83.23          & 96.26          & 91.26          & 87.84                    \\ \cline{3-8} 
                                                                                          &                                                                           & Deep encoder                                                                         & 81.92          & 89.82          & 94.25          & 91.65          & 89.41                    \\ \cline{2-8} 
                                                                                          & Magnitude                                                                 & Deep encoder                                                                         & {\ul 84.10}    & {\ul 90.02}    & {\ul 97.95}    & {\ul 92.07}    & {\ul 91.03}              \\ \hline
\multirow{3}{*}{Proposed}                                                                 & Magnitude                                                                 & Deep encoder                                                                         & 86.75          & \textbf{96.72} & 97.62          & 94.37          & 93.86                    \\ \cline{2-8} 
                                                                                          & \multirow{2}{*}{Complex}                                                  & Deep encoder                                                                         & 87.80          & 95.62          & 98.02          & 94.17          & 93.90                    \\ \cline{3-8} 
                                                                                          &                                                                           & \textbf{\begin{tabular}[c]{@{}c@{}}Deep encoder\\ with attention block\end{tabular}} & \textbf{89.55} & 96.40          & \textbf{98.75} & \textbf{96.87} & \textbf{95.39}           \\ \hline
\end{tabular}
\label{auc table}
\vspace*{-2pt}
\end{table*}

\subsection{Training setup}
For training, spectra are extracted from raw signals with a 64ms analysis frame window and 32ms overlap. We obtain 513-dimensional spectrum features with a 1,024 point fast Fourier transform. We concatenate 2 seconds worth of consecutive input segments (64 frames) from a machine sound by randomly choosing start points.
For the log-mel spectrograms, we compute 128-dimensional mel filterbanks. We used the Adam optimizer \cite{adam} with a learning rate of 0.0001 for training the model. In the testing phase, only softmax values for the combined magnitude and complex features (output $T$ in Fig. \ref{self_supervised_complex_network}) are used to calculate anomaly scores.

\begin{figure}[t]
\centerline{\includegraphics[scale=0.28]{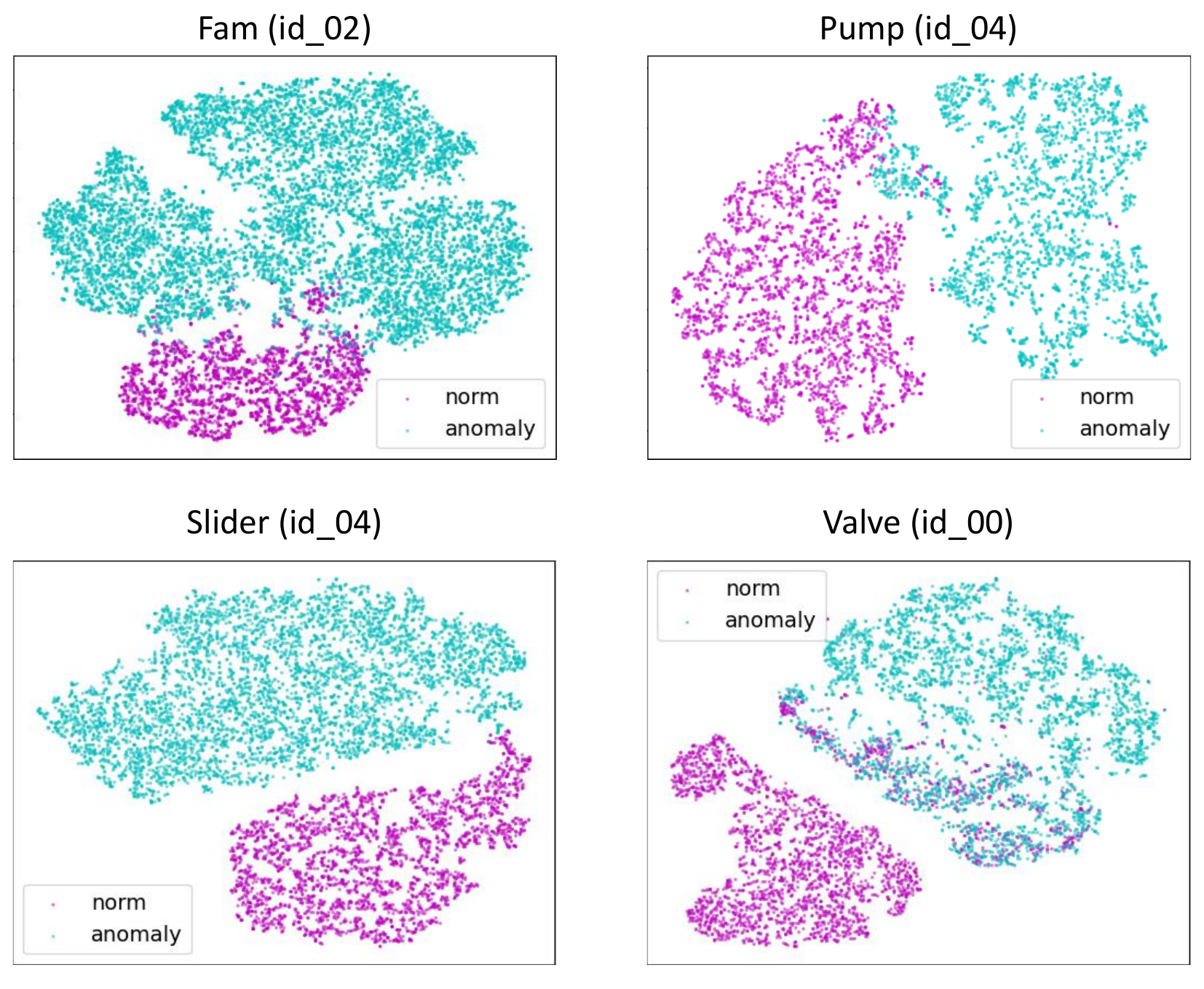}}
\caption{t-SNE plot of each machine sound}
\label{tsne}
\end{figure}

\subsection{Experimental results and analysis}

We evaluate our proposed model using AUC scores for each machine sound type. We also perform an ablation study to check the impact of each of our model's components.
The method in Section \ref{sec:baseline} is used as a baseline method.
\subsubsection{Network architecture dependency}
We compare the performance of our deep encoder network architecture to that of the baseline that uses Mobilenetv2. 
As shown in the first two rows of TABLE \ref{auc table}, the deep encoder structure shows better performance than Mobilenetv2 for most of the machine types. 
We conjecture that using various kernel sizes allows for better feature extraction than Mobilenetv2, which uses a fixed kernel size of $3 \times 3$.

\subsubsection{Input feature dependency (log-mel vs linear magnitude spectrograms)}
We investigate performance variations depending on the type of input features, i.e. log-mel and linear magnitude spectrograms.
Although mel-filterbanks, which correspond to the human auditory system, can reduce the model size, they may also lose key information for anomaly detection. Consequently, linear magnitude spectrum inputs show much improved AUC performance for all machine types compared to log-mel spectrogram inputs, as shown in the second and third system setups in Table \ref{auc table}. 
The underlined numbers represent the best results among the self-supervised classification with linear combination augmentation approach [16].

\subsubsection{Self-supervision task and effect of attention module}
Next, we perform experiments to evaluate the performance variations depending on the type of self-supervised classification tasks.
Compared with the results of the Giri \textit{et al.} [16] given in Table \ref{auc table}, the proposed approach improves AUC scores for most of the machine types with the magnitude spectrum input. Especially for the \textit{pump} type, the proposed approach shows 6.72\% AUC improvement compared to the result of [16]. 
When we only change the input features to complex spectrum, its performance varies by the type of machine sounds, better for the \textit{fan} type but worse for the \textit{pump} type.
However, when we additionally apply an attention block to the complex network, we can obtain the best results among all the models, with an average AUC score of 95.39\%. This specification especially shows the best AUC score for the \textit{fan}, \textit{slider} and \textit{valve} types, achieving 89.55\%, 98.75\% and 96.87\%, respectively. For the \textit{pump} type, however, there is no big difference with the model using magnitude spectrums. The reason for this can be found in Fig. \ref{rainbow}; since there are no remarkable phase continuity differences for \textit{pump} sounds, the impact of phase components to anomaly detection is not significant.


Fig. \ref{tsne} shows t-SNE plots for each machine type to qualitatively show how well our model performs. 
To plot these figures, we used 100 normal samples and 359, 100, 89, 118 anomalous samples for \textit{fan}, \textit{pump}, \textit{slider}, and \textit{valve}, respectively. Perplexity, number of iterations, and learning rate were set to 30, 200, and 1000. 
These plots indicate that normal and anomalous samples are well-clustered by representations from the proposed classification model.

\section{Conclusions}
\label{sec:conclusion}

In this paper, we proposed a self-supervised classification method for machine sound anomaly detection. In consideration of the importance of phase continuity for detecting anomalous sounds, our proposed model introduces a deep complex network with an attention mechanism that controls the relative importance of the magnitude and phase components of the input signal.
Experimental results with the MIMII dataset showed that the AUC scores of the proposed model are much higher than those of the conventional ones.
Our proposed model can be regarded as a generalized anomaly detection system that is effective for all types of machine sounds because we do not need to change the network architecture depending on input signal characteristics.


\section{Acknowledgements}
\label{sec:acknowledgements}

This work was partly supported by the Institute of Information and communications Technology Planning and Evaluation (IITP) grant funded by the Korean government (MIST) (No. 2020-0-01361) and LG Display Co.

\clearpage

\bibliographystyle{IEEEtran}
\bibliography{ref}

\begin{thebibliography}{10}
\providecommand{\url}[1]{#1}
\csname url@samestyle\endcsname
\providecommand{\newblock}{\relax}
\providecommand{\bibinfo}[2]{#2}
\providecommand{\BIBentrySTDinterwordspacing}{\spaceskip=0pt\relax}
\providecommand{\BIBentryALTinterwordstretchfactor}{4}
\providecommand{\BIBentryALTinterwordspacing}{\spaceskip=\fontdimen2\font plus
\BIBentryALTinterwordstretchfactor\fontdimen3\font minus \fontdimen4\font\relax}
\providecommand{\BIBforeignlanguage}[2]{{%
\expandafter\ifx\csname l@#1\endcsname\relax
\typeout{** WARNING: IEEEtran.bst: No hyphenation pattern has been}%
\typeout{** loaded for the language `#1'. Using the pattern for}%
\typeout{** the default language instead.}%
\else
\language=\csname l@#1\endcsname
\fi
#2}}
\providecommand{\BIBdecl}{\relax}
\BIBdecl

\bibitem{unsupervised1}
S.~Ahmad, A.~Lavin, S.~Purdy, and Z.~Agha, ``Unsupervised real-time anomaly detection for streaming data,'' \emph{Neurocomputing}, vol. 262, pp. 134--147, 2017.

\bibitem{unsupervised2}
K.~Leung and C.~Leckie, ``Unsupervised anomaly detection in network intrusion detection using clusters,'' in \emph{Proceedings of the Twenty-eighth Australasian conference on Computer Science-Volume 38}, 2005, pp. 333--342.

\bibitem{unsupervised3}
J.~Inoue, Y.~Yamagata, Y.~Chen, C.~M. Poskitt, and J.~Sun, ``Anomaly detection for a water treatment system using unsupervised machine learning,'' in \emph{2017 IEEE International Conference on Data Mining Workshops (ICDMW)}.\hskip 1em plus 0.5em minus 0.4em\relax IEEE, 2017, pp. 1058--1065.

\bibitem{unsupervsied_ex}
M.~Goldstein and S.~Uchida, ``A comparative evaluation of unsupervised anomaly detection algorithms for multivariate data,'' \emph{PloS one}, vol.~11, no.~4, p. e0152173, 2016.

\bibitem{category}
L.~Bergman and Y.~Hoshen, ``Classification-based anomaly detection for general data,'' \emph{arXiv preprint arXiv:2005.02359}, 2020.

\bibitem{autoencoder1}
E.~Marchi, F.~Vesperini, F.~Eyben, S.~Squartini, and B.~Schuller, ``A novel approach for automatic acoustic novelty detection using a denoising autoencoder with bidirectional lstm neural networks,'' in \emph{Proceedings 40th IEEE International Conference on Acoustics, Speech, and Signal Processing, ICASSP 2015}, 2015, pp. 5--pages.

\bibitem{autoencoder2}
S.~Amiriparian, M.~Freitag, N.~Cummins, and B.~Schuller, \emph{Sequence to sequence autoencoders for unsupervised representation learning from audio}.\hskip 1em plus 0.5em minus 0.4em\relax Universit{\"a}t Augsburg, 2017.

\bibitem{VAE}
M.~S. Kim, J.~P. Yun, S.~Lee, and P.~Park, ``Unsupervised anomaly detection of lm guide using variational autoencoder,'' in \emph{2019 11th International Symposium on Advanced Topics in Electrical Engineering (ATEE)}.\hskip 1em plus 0.5em minus 0.4em\relax IEEE, 2019, pp. 1--5.

\bibitem{autoencoder_gan}
E.~Marchi, F.~Vesperini, S.~Squartini, and B.~Schuller, ``Deep recurrent neural network-based autoencoders for acoustic novelty detection,'' \emph{Computational intelligence and neuroscience}, vol. 2017, 2017.

\bibitem{GAN}
I.~J. Goodfellow, J.~Pouget-Abadie, M.~Mirza, B.~Xu, D.~Warde-Farley, S.~Ozair, A.~Courville, and Y.~Bengio, ``Generative adversarial networks,'' \emph{arXiv preprint arXiv:1406.2661}, 2014.

\bibitem{GMM}
D.~A. Reynolds, ``Gaussian mixture models.'' \emph{Encyclopedia of biometrics}, vol. 741, pp. 659--663, 2009.

\bibitem{GMM1}
B.~Zong, Q.~Song, M.~R. Min, W.~Cheng, C.~Lumezanu, D.~Cho, and H.~Chen, ``Deep autoencoding gaussian mixture model for unsupervised anomaly detection,'' in \emph{International Conference on Learning Representations}, 2018.

\bibitem{GMM2}
H.~Fan, F.~Zhang, R.~Wang, L.~Xi, and Z.~Li, ``Correlation-aware deep generative model for unsupervised anomaly detection,'' in \emph{Pacific-Asia Conference on Knowledge Discovery and Data Mining}.\hskip 1em plus 0.5em minus 0.4em\relax Springer, 2020, pp. 688--700.

\bibitem{OC-SVM}
B.~Sch{\"o}lkopf, R.~C. Williamson, A.~J. Smola, J.~Shawe-Taylor, J.~C. Platt \emph{et~al.}, ``Support vector method for novelty detection.'' in \emph{NIPS}, vol.~12.\hskip 1em plus 0.5em minus 0.4em\relax Citeseer, 1999, pp. 582--588.

\bibitem{dcase2020_outlier-exposer}
P.~Primus, ``Reframing unsupervised machine condition monitoring as a supervised classification task with outlier-exposed classifiers,'' DCASE2020 Challenge, Tech. Rep, Tech. Rep., 2020.

\bibitem{dcase2020_giri}
R.~Giri, S.~V. Tenneti, F.~Cheng, K.~Helwani, U.~Isik, and A.~Krishnaswamy, ``Unsupervised anomalous sound detection using self-supervised classification and group masked autoencoder for density estimation,'' Tech. report in DCASE2020 Challenge Task, Tech. Rep., 2020.

\bibitem{dcase2020_classification_confidence}
T.~Inoue, P.~Vinayavekhin, S.~Morikuni, S.~Wang, T.~H. Trong, D.~Wood, M.~Tatsubori, and R.~Tachibana, ``Detection of anomalous sounds for machine condition monitoring using classification confidence,'' Tech. report in DCASE2020 Challenge Task, Tech. Rep., 2020.

\bibitem{MADE}
M.~Germain, K.~Gregor, I.~Murray, and H.~Larochelle, ``Made: Masked autoencoder for distribution estimation,'' in \emph{International Conference on Machine Learning}.\hskip 1em plus 0.5em minus 0.4em\relax PMLR, 2015, pp. 881--889.

\bibitem{minh_multichannel}
M.~T. Ho, J.~Lee, B.-K. Lee, D.~H. Yi, and H.-G. Kang, ``A cross-channel attention-based wave-u-net for multi-channel speech enhancement,'' in \emph{Proceedings of Interspeech}, 2020, pp. 4049--4053.

\bibitem{importance_phase}
K.~Paliwal, K.~W{\'o}jcicki, and B.~Shannon, ``The importance of phase in speech enhancement,'' \emph{speech communication}, vol.~53, no.~4, pp. 465--494, 2011.

\bibitem{dcunet}
H.-S. Choi, J.-H. Kim, J.~Huh, A.~Kim, J.-W. Ha, and K.~Lee, ``Phase-aware speech enhancement with deep complex u-net,'' in \emph{International Conference on Learning Representations}, 2018.

\bibitem{dcase2020_recognition}
J.~A. Lopez, H.~Lu, P.~Lopez-Meyer, L.~Nachman, G.~Stemmer, and J.~Huang, ``A speaker recognition approach to anomaly detection,'' Tech. report in DCASE2020 Challenge Task, Tech. Rep., 2020.

\bibitem{mimii}
H.~Purohit, R.~Tanabe, K.~Ichige, T.~Endo, Y.~Nikaido, K.~Suefusa, and Y.~Kawaguchi, ``Mimii dataset: Sound dataset for malfunctioning industrial machine investigation and inspection,'' \emph{arXiv preprint arXiv:1909.09347}, 2019.

\bibitem{dcase2020_task2}
Y.~Koizumi, Y.~Kawaguchi, K.~Imoto, T.~Nakamura, Y.~Nikaido, R.~Tanabe, H.~Purohit, K.~Suefusa, T.~Endo, M.~Yasuda \emph{et~al.}, ``Description and discussion on dcase2020 challenge task2: Unsupervised anomalous sound detection for machine condition monitoring,'' \emph{arXiv preprint arXiv:2006.05822}, 2020.

\bibitem{mel}
S.~Umesh, L.~Cohen, and D.~Nelson, ``Frequency warping and the mel scale,'' \emph{IEEE Signal Processing Letters}, vol.~9, no.~3, pp. 104--107, 2002.

\bibitem{rainbowgram}
J.~Engel, C.~Resnick, A.~Roberts, S.~Dieleman, M.~Norouzi, D.~Eck, and K.~Simonyan, ``Neural audio synthesis of musical notes with wavenet autoencoders,'' in \emph{International Conference on Machine Learning}.\hskip 1em plus 0.5em minus 0.4em\relax PMLR, 2017, pp. 1068--1077.

\bibitem{CQT}
J.~C. Brown, ``Calculation of a constant q spectral transform,'' \emph{The Journal of the Acoustical Society of America}, vol.~89, no.~1, pp. 425--434, 1991.

\bibitem{mobilenetv2}
M.~Sandler, A.~Howard, M.~Zhu, A.~Zhmoginov, and L.-C. Chen, ``Mobilenetv2: Inverted residuals and linear bottlenecks,'' in \emph{Proceedings of the IEEE conference on computer vision and pattern recognition}, 2018, pp. 4510--4520.

\bibitem{dccrn}
Y.~Hu, Y.~Liu, S.~Lv, M.~Xing, S.~Zhang, Y.~Fu, J.~Wu, B.~Zhang, and L.~Xie, ``Dccrn: Deep complex convolution recurrent network for phase-aware speech enhancement,'' \emph{arXiv preprint arXiv:2008.00264}, 2020.

\bibitem{adam}
D.~P. Kingma and J.~Ba, ``Adam: A method for stochastic optimization,'' \emph{arXiv preprint arXiv:1412.6980}, 2014.

\end{thebibliography}

\end{document}